\documentclass[a4paper,11pt]{article}

\usepackage{jheppub} 

\usepackage[T1]{fontenc} 
\title{Holographic equipartition from first order action}
\author{Jingbo Wang,}
\affiliation{College of Physics and Electronic Engineering, Xinyang Normal University, Xinyang, 464000, P. R. China}
\emailAdd{shuijing@mail.bnu.edu.cn}

\abstract{In this paper, we get the holographic equipartition form the first order formulism, that is, the connection and its conjugate momentum are considered to be the canonical variables. The final results have similar structure as those from the metric formulism.}

\begin{document}

\maketitle
\flushbottom

\section{Introduction}
The so-called "Emergent Gravity Paradigm" \cite{egp1,egp2,egp3,egp4} developed by T. Padmanabhan and his collaborators suggest that gravity is an emergent phenomenon, and its field equations have the same status as the equations of elasticity or fluid mechanics. An important result is that the evolution of geometry can be interpreted in
thermodynamic terms \cite{padholo1,padholo2}
\begin{equation}\label{1}
    \int_V \frac{d^3 x}{8\pi G}\sqrt{h}u_a g^{ij} L_\xi N^a_{ij}=\frac{1}{2} T_{avg}(N_{sur}-N_{bulk}),
\end{equation}
where
\begin{equation}\label{2}
    N_{sur}=\int_{\partial V}\frac{\sqrt{\sigma}d^2 x}{G};\quad N_{bulk}=\frac{E}{(1/2) T_{avg}};\quad T_{avg}=\frac{1}{A_{sur}}\int_{\partial V} \sqrt{\sigma}d^2 x \frac{N a}{2\pi}
\end{equation}
are the degrees of freedom in the surface and bulk of a 3-dimensional region $V$ and
$T_{avg}$ is the average Davies-Unruh temperature \cite{dut1,dut2} of the boundary. The $h_{ab}$ is
the induced metric on the $t = constant$ surface, $N^a_{ij}=-\Gamma^a_{ij}+\frac{1}{2}(\delta^a_i \Gamma^d_{dj}+\delta^a_j \Gamma^d_{id})$, and $\xi_a = N u_a$ is the proper-time evolution vector corresponding to observers moving
with four-velocity $u_a=-N (dt)_a$, and $E$ is the Komar energy \cite{komar1,komar2}.  The time evolution of the metric in a region
(described by the left hand side), arises because $N_{sur}\neq N_{bulk}$ . In any static spacetime, on the
other hand, $L_\xi(\cdots)=0$, leading to "holographic equipartition" \cite{padholo3,padholo4,padholo5}: $N_{sur}=N_{bulk}$.

In this paper, we will derive the similar expression from the first order formulism, that is the Palatini action. The basic variables are the connection and co-tetrad form fields. This action is closely related to loop quantum gravity \cite{loop1,loop2,loop3,loop4}. Another advantage is that, when considering a black hole, the boundary degrees of freedom can be described by a SO$(1,1)$ BF theory \cite{wmz,wh1,wh2,whl,wh3}. In the following section we set $8\pi G=1$.
\section{The holographic equipartition from first order formulism}
Our starting point is the first order Palatini action coupled with matters,
\begin{equation}\label{3}\begin{split}
    S[e,A]=-\frac{1}{4}\int_{\mathcal{M}} \varepsilon_{IJKL} e^I\wedge e^J \wedge F(A)^{KL}+\frac{1}{2}\int_{\mathcal{M}} e^I \wedge T_I\\
    = -\frac{1}{2}\int_{\mathcal{M}} \Sigma_{IJ} \wedge F(A)^{KL}+\frac{1}{2}\int_{\mathcal{M}} e^I \wedge T_I,
\end{split}\end{equation}
where $e^I$ are covielbein 1-form fields,
\begin{equation}\label{4}
    \Sigma_{IJ}=\frac{1}{2}\varepsilon_{IJKL}e^K \wedge e^L
\end{equation}
 2-form fields, $A^{IJ}$ the $SO(3,1)$ connection 1-form, $F_{IJ}$ the curvature 2-form
of $A^{IJ}$, $T_I$ the energy-momentum 3-form and $I,J$ indices of the Lie algebra of $\frak{so}(3,1)$.
From the above action one can get the field equations,
\begin{equation}\label{5}
\varepsilon_{IJKL}e^J\wedge F(A)^{KL}=T_I,\quad {\rm d}_A \Sigma_{IJ}:={\rm d}\Sigma_{IJ}-A_I^{\ K}\wedge\Sigma_{KJ}-A_J^{\ K}\wedge\Sigma_{IK}=0.
\end{equation}

The dynamical variables can be chosen as the connection $A^{IJ}$ and its conjugate momentum $\Sigma_{IJ}$. Similar to Eq.(\ref{1}), we want to calculate the following expression for the first order formula
\[ \int_V \Sigma_{IJ} \wedge L_\xi A^{IJ}.\]
Due to the Cartan formula $L_\xi={\rm d} i_\xi+i_\xi {\rm d}$, where $i_X$ is the interior product, we get
\begin{equation}\label{6}\begin{split}
  \int_V \Sigma_{IJ} \wedge L_\xi A^{IJ}=\int_V \Sigma_{IJ} \wedge [ {\rm d}(i_\xi A^{IJ})+i_\xi {\rm d}A^{IJ}]\\
  =\int_V {\rm d}(\Sigma_{IJ} \wedge i_\xi A^{IJ})-{\rm d}\Sigma_{IJ} \wedge i_\xi A^{IJ}+\Sigma_{IJ} \wedge i_\xi {\rm d}A^{IJ}\\
   =\int_V \Sigma_{IJ} \wedge i_\xi {\rm d}A^{IJ}-{\rm d}\Sigma_{IJ} \wedge i_\xi A^{IJ}+\int_{\partial V} \Sigma_{IJ} \wedge i_\xi A^{IJ}.
\end{split}\end{equation}

\subsection{The bulk term}
First let us consider the bulk term. We choose the time gauge as
\begin{equation}\label{7}
    e^0_a=u_a=N\xi_a=-N(dt)_a\Rightarrow e_0^a=-u^a=-\xi^a/N,
\end{equation}
so on the hypersurface $V:t=constant$, we have
\begin{equation}\label{8}
    e^0\triangleq 0.
\end{equation}
(We denote equalities on $V$ by the symbol $\triangleq$.) Due to the definition (\ref{4}), the only non-zero $\Sigma$ are $\Sigma_{0i} (i=1,2,3)$. The bulk term is \begin{equation}\label{9}\begin{split}
     2 \int_V \Sigma_{0i} \wedge i_\xi {\rm d}A^{0i}-{\rm d}\Sigma_{0i} \wedge i_\xi A^{0i}=2 \int_V \Sigma_{0i} \wedge i_\xi (F^{0i}-A^0_{\ k} \wedge A^{ki})-i_\xi A^{0i} (A^k_{\ i}\wedge \Sigma_{0k}) \\
     =2 \int_V \Sigma_{0i} \wedge i_\xi F^{0i}+i_\xi A^{ki} (A_{\ k}^0\wedge \Sigma_{0i}),
\end{split}\end{equation}
where the field equations (\ref{5}) are used.

Since $\xi^a=-N e_0^a$,
\begin{equation}\label{10}\begin{split}
   2 \int_V \Sigma_{0i} \wedge i_\xi F^{0i}=2 \int_V i_\xi (\Sigma_{0i}\wedge F^{0i})= 2 \int_V i_\xi \frac{1}{2}(e^i \wedge T_i-e^0 \wedge T_0)\\
   =\int_V N (T_0 -e^i \wedge <e_0,T_i>),
\end{split}\end{equation}
where $<e_0,T_i>=i_{e_0} T_i$.

The second term can be written as
\begin{equation}\label{11}\begin{split}
         2 \int_V i_\xi A^{ki} (A_{\ k}^0\wedge \Sigma_{0i})=2 \int_V <-Ne_0, A^{ki}><e_i, A_{\ k}^0> e^1\wedge e^2\wedge e^3\\
         =\int_V -N <e_0, A^{ki}>(<e_i, A_{\ k}^0>-<e_k, A_{\ i}^0>) e^1\wedge e^2\wedge e^3.
\end{split}\end{equation}
Since the connection 1-form fields are given by
\begin{equation}\label{11a}
    A_I^{\ J}=e_I^c \nabla_a e^J_c dx^a,
\end{equation}
one can get
 \begin{equation}\label{12}\begin{split}
   <e_i, A_{\ k}^0>-<e_k, A_{\ i}^0>=-(e^a_i e^c_k \nabla_a e^0_c-e^a_k e^c_i \nabla_a e^0_c)=-(e^a_i e^c_k \nabla_a e^0_c-e^c_k e^a_i \nabla_c e^0_a)\\
   =-e^a_i e^c_k (\nabla_a e^0_c-\nabla_c e^0_a)=-e^a_i e^c_k ({\rm d} e^0)_{ac},
\end{split}\end{equation}
since $e^0_a=-N (dt)_a$, one can get $d e^0=-d N\wedge dt, e_i^0=0$, so this term is 0.

The bulk term can be written as
\begin{equation}\label{13}\begin{split}
  \int_V \Sigma_{IJ} \wedge i_\xi {\rm d}A^{IJ}-{\rm d}\Sigma_{IJ} \wedge i_\xi A^{IJ}=\int_V N (T_0 -e^i \wedge <e^0,T_i>)\equiv E,
\end{split}\end{equation}
where $E$ can be considered as the energy of the system.
\subsection{The boundary term}
Next we consider the boundary term. We choose the boundary $\partial V$ to be $N=constant$ surface within $t=constant$ surface \cite{padholo1}.
Define the acceleration vector $a^c=u^b \nabla_b u^c$, which satisfy $a^b u_b=0$, so we can choose $e_c^1=\frac{a_c}{\sqrt{\|a^c a_c\|}}=\frac{a_c}{a}$.
On the other hand,
\begin{equation}\label{15}
    a_c=h_c^b \nabla_b N/N \circeq 0,
\end{equation}
where $\circeq$ means the equalities on $\partial V$, so we have
\begin{equation}\label{16}
    e^1 \circeq 0.
\end{equation}
Thus the no zero covielbein on $\partial V$ are $e^2,e^3$. The boundary term is
\begin{equation}\label{14}\begin{split}
    \int_{\partial V} \Sigma_{IJ} \wedge i_\xi A^{IJ}=2\int_{\partial V} \Sigma_{01} \wedge i_\xi A^{01}
    =2\int_{\partial V} <-N e_0, A^{01}> e^2 \wedge e^3\\
    =2\int_{\partial V} -N (e^a_0 e^c_1 \nabla_a e^0_c) e^2 \wedge e^3=2\int_{\partial V} -N \sqrt{\|a^c a_c\|} e^2 \wedge e^3
    =2\int_{\partial V} -N a e^2 \wedge e^3=-2\int_{\partial V} N a \sqrt{\sigma}d^2 x.
\end{split}\end{equation}

So the final result is
\begin{equation}\label{15}\begin{split}
  \int_V \Sigma_{IJ} \wedge L_\xi A^{IJ}=\int_{\partial V} -2 N a \sqrt{\sigma}d^2 x+\int_V N (T_0 -e^i \wedge <e^0,T_i>).
\end{split}\end{equation}

Define the surface and bulk degrees of freedom,
\begin{equation}\label{16}
    N_{sur}=\int_{\partial V}\frac{\sqrt{\sigma}d^2 x}{G}=8\pi\int_{\partial V}\sqrt{\sigma}d^2 x;\quad N_{bulk}=\frac{E}{(1/2) T_{avg}},
\end{equation}
then the time evolution of spacetime can be written as
\begin{equation}\label{17}\begin{split}
  -\int_V \Sigma_{IJ} \wedge L_\xi A^{IJ}=\frac{1}{2} T_{avg} (N_{sur}-N_{bulk}),
\end{split}\end{equation}similar to Eq.(\ref{1}).
\section{Conclusion}
In this paper, the main result is the Eq.(\ref{17}), which states that the time evolution of the spacetime is driven by the departure from the holographic equipartition. It has the similar structure as that from the metric formulism. In all static spacetime, the time evolution is frozen, so maintain the holographic equipartition.

\acknowledgments
The author would like to thank Dr. Yu Han and Dr. Molin Liu for many help.

\bibliography{holo1}

\providecommand{\href}[2]{#2}\begingroup\raggedright\begin{thebibliography}{10}

\bibitem{egp1}
T.~Padmanabhan, {\it {Emergent Gravity Paradigm: Recent Progress}},  {\em Mod.
  Phys. Lett.} {\bf A30} (2015), no.~03n04 1540007,
  [\href{http://xxx.lanl.gov/abs/1410.6285}{{\tt arXiv:1410.6285}}].

\bibitem{egp2}
T.~Padmanabhan, {\it {Gravity and the thermodynamics of horizons}},  {\em Phys.
  Rept.} {\bf 406} (2005) 49--125,
  [\href{http://xxx.lanl.gov/abs/gr-qc/0311036}{{\tt gr-qc/0311036}}].

\bibitem{egp3}
T.~Padmanabhan, {\it {Thermodynamical Aspects of Gravity: New insights}},  {\em
  Rept. Prog. Phys.} {\bf 73} (2010) 046901,
  [\href{http://xxx.lanl.gov/abs/0911.5004}{{\tt arXiv:0911.5004}}].

\bibitem{egp4}
S.~Chakraborty and T.~Padmanabhan, {\it {Thermodynamical interpretation of the
  geometrical variables associated with null surfaces}},  {\em Phys. Rev.} {\bf
  D92} (2015), no.~10 104011, [\href{http://xxx.lanl.gov/abs/1508.0406}{{\tt
  arXiv:1508.0406}}].

\bibitem{padholo1}
T.~Padmanabhan, {\it {General Relativity from a Thermodynamic Perspective}},
  {\em Gen. Rel. Grav.} {\bf 46} (2014) 1673,
  [\href{http://xxx.lanl.gov/abs/1312.3253}{{\tt arXiv:1312.3253}}].

\bibitem{padholo2}
T.~Padmanabhan, {\it {What drives the time evolution of the spacetime
  geometry?}},  {\em Int. J. Mod. Phys.} {\bf D23} (2014), no.~12 1441003,
  [\href{http://xxx.lanl.gov/abs/1405.5535}{{\tt arXiv:1405.5535}}].

\bibitem{dut1}
P.~C.~W. Davies, {\it {Scalar particle production in Schwarzschild and Rindler
  metrics}},  {\em J. Phys.} {\bf A8} (1975) 609--616.

\bibitem{dut2}
W.~G. Unruh, {\it {Notes on black hole evaporation}},  {\em Phys. Rev.} {\bf
  D14} (1976) 870.

\bibitem{komar1}
A.~Komar, {\it {Covariant conservation laws in general relativity}},  {\em
  Phys. Rev.} {\bf 113} (1959) 934--936.

\bibitem{komar2}
R.~M. Wald, {\em {General Relativity}}.
\newblock Chicago, Usa: Univ. Pr., 1984.

\bibitem{padholo3}
T.~Padmanabhan, {\it {Gravitational entropy of static space-times and
  microscopic density of states}},  {\em Class. Quant. Grav.} {\bf 21} (2004)
  4485--4494, [\href{http://xxx.lanl.gov/abs/gr-qc/0308070}{{\tt
  gr-qc/0308070}}].

\bibitem{padholo4}
T.~Padmanabhan, {\it {Equipartition of energy in the horizon degrees of freedom
  and the emergence of gravity}},  {\em Mod. Phys. Lett.} {\bf A25} (2010)
  1129--1136, [\href{http://xxx.lanl.gov/abs/0912.3165}{{\tt
  arXiv:0912.3165}}].

\bibitem{padholo5}
T.~Padmanabhan, {\it {Surface Density of Spacetime Degrees of Freedom from
  Equipartition Law in theories of Gravity}},  {\em Phys. Rev.} {\bf D81}
  (2010) 124040, [\href{http://xxx.lanl.gov/abs/1003.5665}{{\tt
  arXiv:1003.5665}}].

\bibitem{loop1}
C.~Rovelli, {\em Quantum Gravity}.
\newblock Cambridge Monographs on Mathematical Physics. Cambridge University
  Press, 2004.

\bibitem{loop2}
T.~Thiemann, {\em Modern Canonical Quantum General Relativity}.
\newblock Cambridge Monographs on Mathematical Physics. Cambridge University
  Press, 2008.

\bibitem{loop3}
A.~Ashtekar and J.~Lewandowski, {\it Background independent quantum giravity: a
  status report},  {\em Classical and Quantum Gravity} {\bf 21} (2004), no.~15
  R53--R152.

\bibitem{loop4}
M.~X. Han, Y.~G. Ma, and W.~M. Huang, {\it Fundamental structure of loop
  quantum gravity},  {\em International Journal of Modern Physics D} {\bf 16}
  (2007), no.~9 1397--1474.

\bibitem{wmz}
J.~B. Wang, Y.~G. Ma, and X.~A. Zhao, {\it {BF theory explanation of the
  entropy for non-rotating isolated horizons}},  {\em Phys Rev} {\bf D89}
  (2014) 084065, [\href{http://xxx.lanl.gov/abs/1401.2967}{{\tt
  arXiv:1401.2967}}].

\bibitem{wh1}
J.~B. Wang and C.~G. Huang, {\it {The entropy of higher dimensional nonrotating
  isolated horizons from loop quantum gravity}},  {\em Class Quant Grav} {\bf
  32} (2015) 035026.

\bibitem{wh2}
J.~B. Wang and C.~G. Huang, {\it {BF theory explanation of the entropy for
  rotating isolated horizons}},  {\em Int. J. Mod. Phys. D} {\bf 25} (2016)
  1650100.

\bibitem{whl}
J.~B. Wang, C.~G. Huang, and L.~Li, {\it {The Entropy of nonrotating isolated
  horizons in Lovelock theory from loop quantum gravity}},  {\em Chinese Phys.
  C} {\bf 40} (2016) 083102.

\bibitem{wh3}
J.~B. Wang and C.~G. Huang, {\it {The entropy of isolated horizons in
  non-minimally coupling scalar field theory from BF theory}},
  \href{http://xxx.lanl.gov/abs/1507.0880}{{\tt arXiv:1507.0880}}.

\end{thebibliography}\endgroup
\end{document}